# Rethinking Polarization in Wurtzite Semiconductors

*Ding Wang[1]\*, Danhao Wang[1]\*, Samuel Yang[1], Zetian Mi[1]\**

[1]Department of Electrical Engineering and Computer Science, University of Michigan, Ann Arbor, 48109, USA.

\*Corresponding author. Email: dinwan@umich.edu; danhaow@umich.edu; ztmi@umich.edu

**Polarization arising from non-centrosymmetric wurtzite lattice underpins the physics and functionality of gallium nitride (GaN) – the most produced semiconductor materials second only to silicon. However, recent direct experimental measurements unveiled remanent polarization of unexpectedly large magnitudes and opposite orientations to traditionally anticipated. This significant discrepancy not only poses a formidable challenge to our existing theoretical paradigms but also accentuates the need for a critical rethinking and methodological refinement to integrate these novel observations with established knowledge, mitigating potential misunderstandings and misconceptions in this rapidly evolving field.**

## Main Text

Over the past three decades, group III-nitrides (GaN, AlN, InN, and their alloys) have been extensively investigated across various fields, including electronics, optoelectronics, piezo-electronics, quantum-photonics, and clean energy. III-nitrides exhibit a stable non-centrosymmetric wurtzite phase. The non-equivalence between the *c*-axis bond and the *c*-components of the other three bonds results in a non-zero electric dipole within the structure, leading to strong spontaneous and piezoelectric polarization along the *c*-direction and plays a critical role in determining the optical and electrical properties of wurtzite nitrides. For example, differences in polarization induce high densities of two-dimensional carrier gases (2DCGs) at the heterostructure interfaces, which have been exploited in modern high-frequency and high-power transistors.[1-3] The polarization discontinuities also give rise to unwanted built-in electric fields in quantum wells. The resulting quantum-confined Stark effect (QCSE) reduces the electron-hole wave-function overlap and thus the efficiency of light emitters. [4,5]

Based on the Modern Theory of Polarization (MTP),[6] the polarization constants have been well-determined in Bernardini *et al.'s* seminal work by using a zinc-blende structure as reference.[7] The effective application of this theoretical framework in quantitatively interpreting and modeling polarization phenomena in nitride heterostructures has garnered widespread recognition and textbook-level acclaim, serving as a quintessential example of successfully bridging theory with experimental observations.

Recently, the polarization in wurtzite nitride semiconductors have been unambiguously measured experimentally by different groups.[8-10] By introducing rare-earth elements into the wurtzite structure, the polarity switching energy barrier is significantly reduced, permitting the first experimental measurement of spontaneous polarization in the wurtzite semiconductors.[8] In an unexpected twist, the observed spontaneous polarization values are consistently an order of magnitude larger and orients in direct opposition to the predictions made by Bernardini *et al*. that have informed the understanding of wurtzite semiconductors for decades. [8-12] These fundamental discrepancies profoundly disrupt our foundational comprehension of polarization in these materials and pose significant challenges on both the theoretical framework and the device development. It is therefore imperative to reevaluate the theoretical framework for polarization phenomena in wurtzite semiconductors.

In this comment, we discuss research efforts in determining the polarization constants of wurtzite semiconductors. Our aims are to draw attention to a critical rethinking of the polarization phenomena in wurtzite semiconductors and to provide some clarification on theories and experiments that are plagued by possible misconceptions.

**Conventional Understanding of Polarization in Group-III Nitrides**

The wurtzite structure is characterized by a hexagonal Bravais lattice with four atoms per unit cell and by three critical parameters: the edge length $a$ of the basal hexagon, the height $c$ of the hexagonal prism, and an internal parameter $u$ expressed as a fraction of $c$, which determines the relative positions of atoms along the $c$-axis. The $c/a$ ratio in the wurtzite structure, being less than the ideal tetrahedral value of $\sqrt{8/3} \approx 1.633$, induces a displacement between cation and anion centers, generating a net dipole moment along the $c$-axis. To attain a mental image of polarization in the wurtzite structure, historically, a tetrahedron has been frequently referenced. When the tetrahedron is ideal ($u = 3/8$), all polarization vectors cancel one another. In the case of III-nitride wurtzite structure ($u > 3/8$), the central atom is not exactly located at the tetrahedral center, leading to a non-zero electric polarization along $c$-axis, known as the spontaneous polarization. Under homogeneous in-plane strain, the sum of the polarization vectors varies, causing a net polarization along the $c$-direction, known as the piezoelectric polarization.

In the 1990s, it became apparent that the more fundamental measure of polarization is the differential polarization, a concept that was further developed by King-Smith, Vanderbilt, and Resta with the advent of the MTP.[6] This approach has been recommended for defining the spontaneous polarization of a material, effectively addressing the challenges of multivalued formal polarization. Specifically, for wurtzite semiconductors, the zincblende structure has been utilized as a reference by Bernardini *et al.* to determine the polarization constants.[7] The effective polarization of the wurtzite structure is conceptualized as a net dipole moment arising from the displacement of the center atom, with the direction of spontaneous polarization oriented opposite to the *c*-axis. Subsequent calculations of polarization constants for III-nitrides have been extensively applied in interpreting and modeling the key phenomena including 2DCGs and internal fields in quantum wells.[5,13]

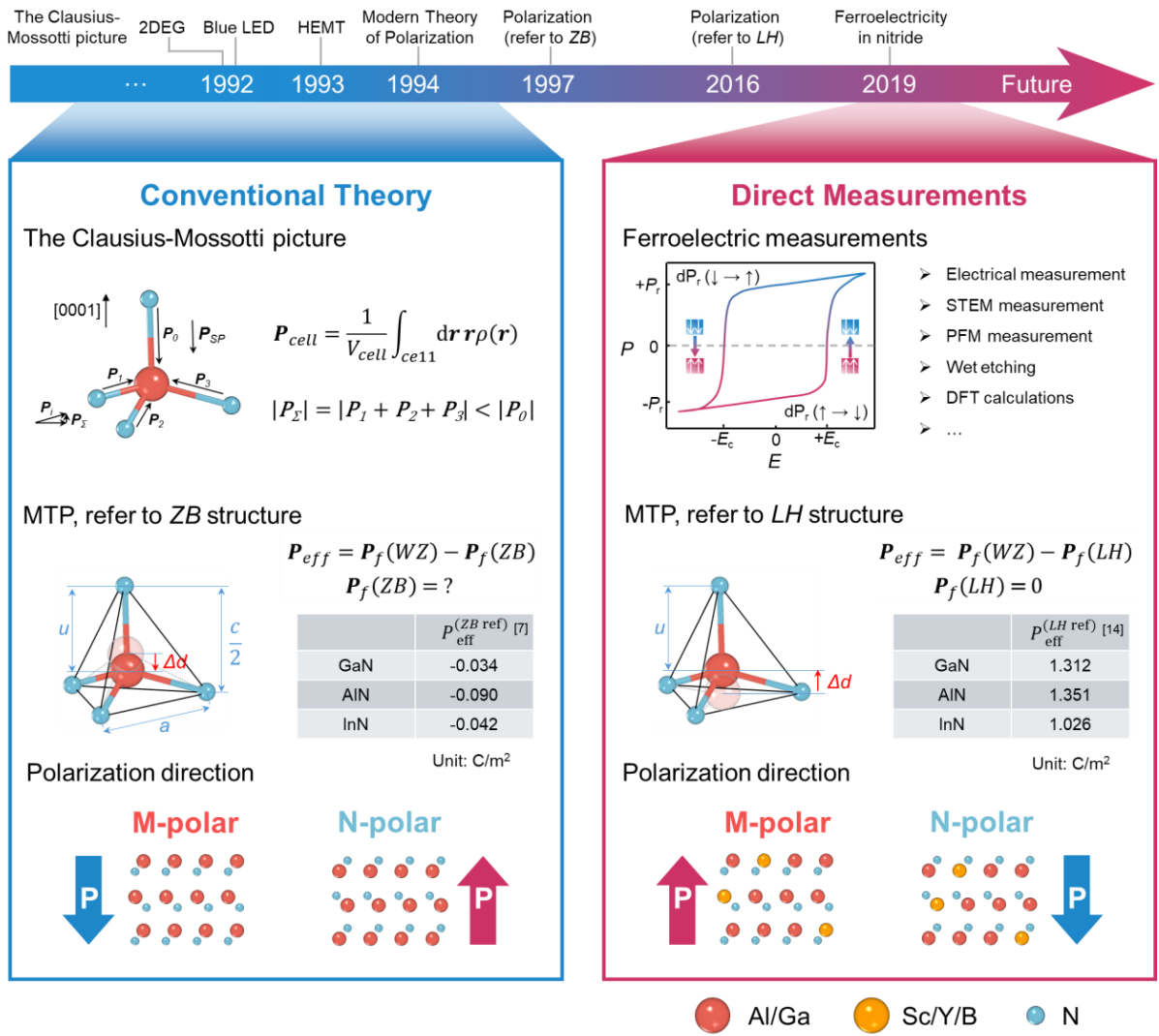

**Fig. 1**. Initially, the polarization was intuitively understood as a cumulative polarization of the bonds in an exemplary tetrahedron divided by the unit volume. Extensive interest was raised in III-nitrides after the demonstration of 2DEGs and HEMTs by Asif Khan *et al.*,[1,2] and the first high-efficiency blue light emitting diodes (LEDs) by Akasaki *et al.*[4] Inspired by the development of the MTP by Resta *et al.*,[6] in 1997, Bernardini *et al* first calculated the polarization constants in group-III nitrides using the Berry phase approach and the zincblende (ZB) structure as reference.[7] The theoretical results suggested a spontaneous polarization pointing towards -*c* direction yet the zero-polarization assumption for zincblende structure remains questionable. In 2016, Dreyer *et al* proposed a layered hexagonal (LH) structure as an alternative reference structure for calculating the polarization constants in wurtzite nitrides.[14] However, due to the lack of experimental proof, the results were merely regarded as magnitude refinements. Very recently, direct experimental measurement of spontaneous polarization using wurtzite ferroelectrics was done by various groups.[8-10] Surprisingly, the measured spontaneous polarization orients oppositely to and is one order larger in magnitude than conventional belief yet correlates well with Dreyer *et al.'s* calculations, and have further been validated by scanning transmission electron microscopy (STEM), piezo-response force microscopy (PFM), wet-etching and Density Function Theory (DFT) calculations.[11,12] These advancements highlight the validity of using layered hexagonal structure as a standard reference and call for a critical rethinking of the polarization in wurtzite semiconductors.

**Direct Measurements of Polarization in Wurtzite Nitride Semiconductors**

While the theoretical framework has proven effective for conventional nitrides, applying the same methodology to predict spontaneous polarization in rare-earth-doped III-nitrides presents significant challenges. The discovery of ferroelectricity in wurtzite nitrides has enabled direct measurements of spontaneous polarization in this material class.[8] However, the breakthrough brought not only excitement but also questions:

(1) The direction of spontaneous polarization determined by macroscopic ferroelectric measurements is oriented exactly opposite to the predictions made by Bernardini *et al.* This discrepancy in lattice polarity orientation relative to measured polarization has been further validated by examining atomic configurations under varying external electric fields.[11,12]

(2) The polarization values measured are an order of magnitude greater than those previously predicted using a zincblende reference structure.

These results underscore a critical rethinking and reevaluation of polarization in wurtzite semiconductors from the ground up, in order to reconcile the fundamental discrepancies between conventional theory and recent experimental observations.

**Layered Hexagonal to the Rescue**

What happened to the zincblende reference structure? Upon scrutinizing the criteria for selecting reference structures within the MTP, one realizes that the common practice of using zincblende structure as reference may not fully align with the stringent requirements, prompting a reevaluation of its suitability as a standard reference for calculating the polarization in wurtzite materials.

**(1) Challenges in precisely defining polarization in individual wurtzite materials**

In the calculations by Bernardini *et al.*, the interface theorem was consistently applied to circumvent the necessity of identifying a complex adiabatic, gap-preserving deformation pathway.[7] If N- and M-polarities can be defined in the zincblende structure based on the orientation of the vertical bond in the [111] direction, similar to that in the wurtzite structure, one would find that the insulating interface constructed between wurtzite and zincblende structures used in their calculations are only based on either N-polar wurtzite to N-polar zincblende, or M-polar wurtzite to M-polar zincblende.[15] Once polarization of one polarity is determined, the polarization for the opposite polarity is simply assumed as inversely equivalent. This, however, causes problems because the reference structures are substantially different while calculating the polarization of opposing polarities. An intuitive physical picture is that, the transition from M-polar zincblende to M-polar wurtzite requires a minimal displacement of the central atom by *(u-3/8)c*. However, the inverted operation of displacing the central atom in M-polar zincblende by -(*u*-3/8)*c* ends up in a structure that deviates from the expected N-polar wurtzite structure. Instead, a much larger displacement of -(1-*u*-3/8)*c* is required. This suggests that utilizing the zincblende structure as a reference can introduce complications in accurately defining the spontaneous polarization even in a single wurtzite material.

**(2) Difficulties in accurately determining polarization at interfaces**

Contrary to prevailing assumptions, the formal polarization of the zincblende structure does not inherently vanish along the [111] direction.[6,14] A closer examination of its ionic contribution reveals that the structure can be visualized as a series of displaced one-dimensional ion chains along the [111] direction, which exhibit zero polarization only for specific unit cell choices. The value of this conditional non-vanishing polarization is highly dependent on the lattice constant of the chosen reference structure, introducing inaccuracies when calculating the difference of formal polarization across interfaces. In other words, when using a zincblende structure reference, variations in the lattice constant across the interface would lead to varying polarization reference values for the materials constituting the interface, which, however, was not considered in conventional Bernardini *et al.*'s theory.

**(3) Layered hexagonal as a standard reference**

An ideal reference structure should satisfy several key criteria. First, it must guarantee that, while deforming the reference structure into the polarization state, the inverted polarization states can be obtained via inverted deforming paths. Second, the reference structure should exhibit no interface charge when an interface is formed between itself and its mirrored counterpart across the plane of interest, thus assuring vanishing formal polarization. Additionally, to enable accurate calculation of net polarization charges at material interfaces using polarization constants, the chosen reference structures for materials constructing the interface should exhibit vanishing polarization, even under stress applied perpendicular to the direction of interest.

By constructing an interface between N-polar and M-polar wurtzite lattices and adhering to the criteria previously outlined, the layered hexagonal structure is unequivocally identified as an ideal reference structure. This approach not only yields calculated magnitudes and orientations of spontaneous polarization that closely match those observed in direct experimental measurements but also effectively resolves the discrepancies highlighted earlier, affirming the validity of this reference choice. [14] This theoretical framework currently finds its acceptance predominantly within the ferroelectric community, yet with only limited acknowledgment in traditional III-nitride research. We highlight the necessity of merging these perspectives to refine and expand the theoretical and empirical understanding of polarization phenomena in all wurtzite semiconductors.

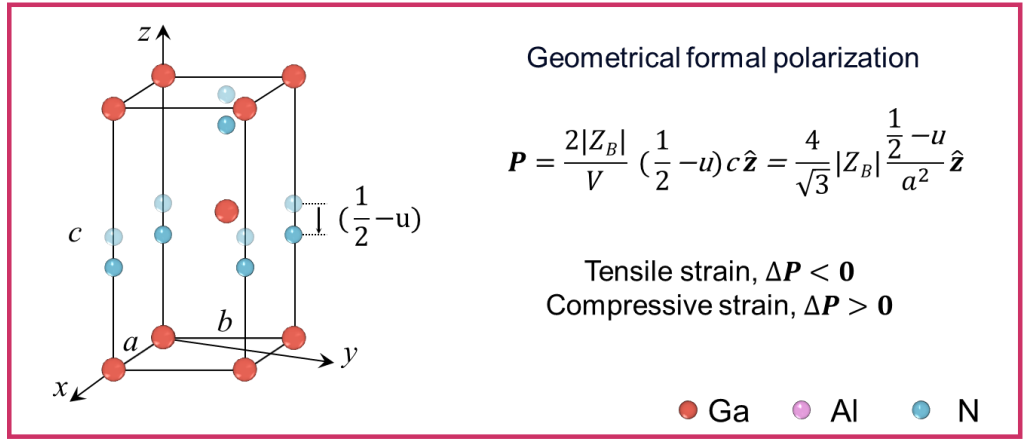

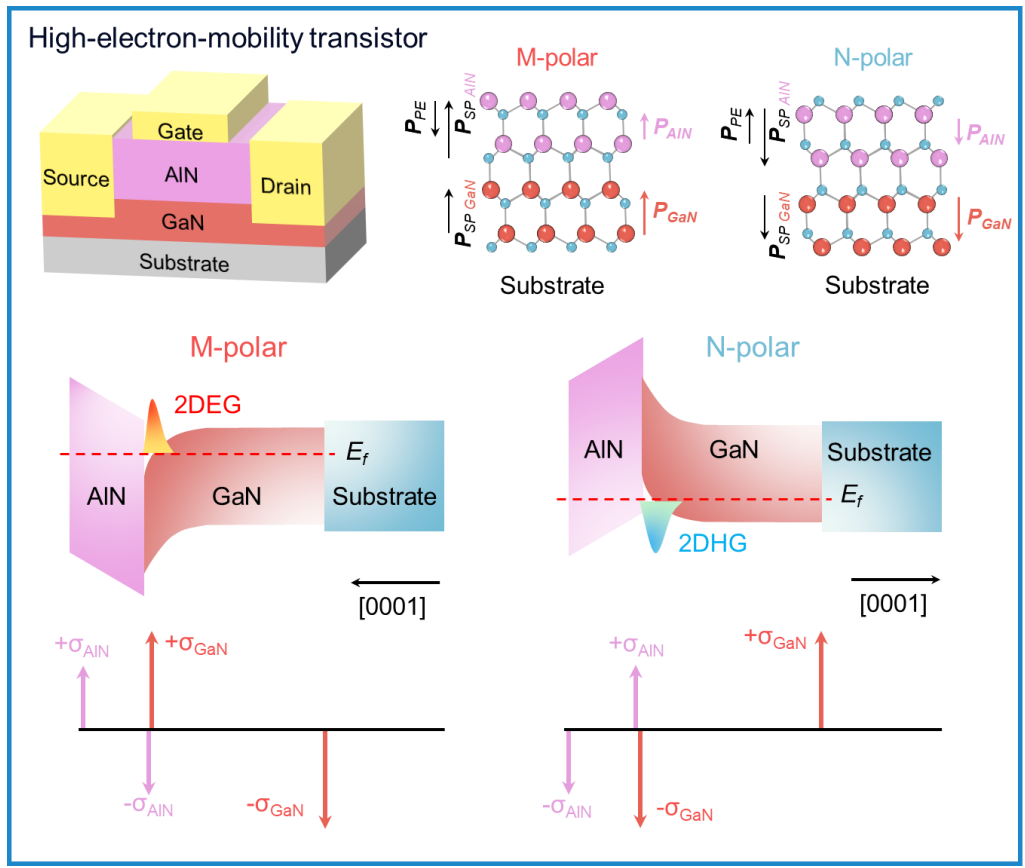

**Fig. 2**. The spontaneous polarization can be estimated by the dipole moment due to a displacement of the nitrogen atoms from a layered hexagonal structure to a wurtzite structure, divided by the volume of the unit cell. $Z_B$ is the Born effective charge. Tensile strain leads to an increase in $u$ and $a$, resulting in reduced polarization, whereas compressive strain results in an increase in polarization. Implementing the new direction of spontaneous polarization and piezoelectric polarization, the 2DEG and 2DHG could be explained as a result of the larger polarization of GaN compared to AlN, contrary to the physical picture used before.

**Rethinking the Polarization in III-nitrides**

Here, we recall the significant conclusions from the experimental and computational results based on direct measurements and the layered hexagonal reference: (1) The magnitude of spontaneous polarization in these materials exceeds previous estimates by an order of magnitude. (2) Contrary to conventional belief, the orientation of spontaneous polarization aligns with $+c$ instead of $-c$ direction. (3) Consequently, piezoelectric polarization due to tensile strain now reduces, rather than enhances, the total polarization. These revelations compel a thorough reevaluation of the conceptual framework underlying the key phenomena of nitride systems, including 2DCGs, internal fields, surface band bending, photocurrents, and polarization doping, to name a few.

To holistically comprehend polarization in wurtzite materials, it becomes imperative to investigate lateral polarity heterostructures and heterostructures that are relaxed, whether partially or fully. Such structures are essential candidates to validate the revised polarization constants. Moreover, the application of non-electrical experimental methods, such as scanning probe microscopy and differential phase contrast imaging, promises to further illuminate the intrinsic polarization characteristics of nitrides.

We note that for both reference structures, the concept of differential polarization is considered, therefore the *proper* piezoelectric constants should be consistent with each other and with piezoelectric measurements. For use in the interpretation or modeling of experiments involving interface charges, to account for the dilution or concentration of "pre-strain" bound charge, *improper* piezoelectric constants should be used, especially for materials with large spontaneous polarization.[16] This part was treated by adding a correction term to the *proper* ones by Dreyer.[14] In a recent study, it is demonstrated that employing the layered hexagonal reference allows polarization constants of ternary wurtzite semiconductors to be precisely and comfortably described using a linear Vegard's law.[17] These advancements highlight the effectiveness of using polarization constants derived from a layered hexagonal reference as physical descriptors for polarization in wurtzite semiconductors.

As a step forward, we attempt to implement the new theoretical and experimental results in understanding one of the most important phenomena in wurtzite nitride semiconductors: the 2DCGs in the AlN/GaN system. According to the new theoretical and experimental results, the direction of spontaneous polarization is in the $+c$ direction.[14] The larger spontaneous polarization of AlN over GaN implies that the sign of the interface charge will reverse if spontaneous polarization is considered alone. Therefore, the tensile stress and corresponding piezoelectric polarization of AlN must exceed the difference in spontaneous polarization and ensure a net positive charge at the interface to support the generation of 2DEGs. This suggests that the formation of the 2DEG, and by extension, the two-dimensional hole gas (2DHG) in N-polar systems, as well as polarization doping via composition grading, may predominantly rely on piezoelectric effects.

While strain is often closely associated with local compositional variance, defects, and relaxation, the 2DEG in nitride systems is remarkably robust. This phenomenon warrants further discussion and offers new avenues for future polarization engineering and device design. A preliminary consideration is that, since the internal parameter $u$ of AlN is greater than that of GaN, the cumulative polarization will always be less than that of GaN when AlN is grown coherently on top of GaN. The same consideration applies for AlGaN barriers. Therefore, regardless of the Al composition, a net positive charge, and thus robust 2DEG is guaranteed. Further calculations are necessary to elucidate this unique phenomenon in more depth.

**Summary**

The recent direct measurement of polarization in wurtzite nitride semiconductors has posed significant challenges for existing theories and underscores an urgent need for critical rethinking and methodological improvements in our basic understanding of polarization phenomena in wurtzite semiconductors. Such a critical rethinking will lay the foundation for the physics, design, and functionality of a broad range of nitride-based electronic, photonic, and piezoelectric devices.

**References**

1  Khan, M. A., Kuznia, J. N., Van Hove, J. M., Pan, N. & Carter, J. *Appl. Phys. Lett.* **60**, 3027-3029 (1992).
2  Asif Khan, M., Bhattarai, A., Kuznia, J. N. & Olson, D. T. *Appl. Phys. Lett.* **63**, 1214-1215 (1993).
3  Mishra, U. K., Parikh, P. & Wu, Y. F. *Proc. IEEE.* **90**, 1022-1031 (2002).
4  Akasaki, I. in *Inst. Phys. Conf. Ser.*  851-856 (1992).
5  Deguchi, T. *et al. Jpn J Appl Phys 2* **38**, L914-L916 (1999).
6  Resta, R. & Vanderbilt, D. *Top Appl Phys* **105**, 31-68 (2007).
7  Bernardini, F., Fiorentini, V. & Vanderbilt, D. *Phys. Rev. B* **56**, 10024-10027 (1997).
8  Fichtner, S., Wolff, N., Lofink, F., Kienle, L. & Wagner, B. *J. Appl. Phys.* **125**, 114103 (2019).
9  Hayden, J. *et al. Phys. Rev. Mater.* **5**, 044412 (2021).
10  Wang, D., Wang, P., Wang, B. Y. & Mi, Z. T. *Appl. Phys. Lett.* **119**, 111902 (2021).
11  Calderon, V. S. *et al. Science* **380**, 1034-1038 (2023).
12  Wang, D. *et al. arXiv preprint arXiv:2312.08645* (2023).
13  Ambacher, O. *et al. J. Appl. Phys.* **87**, 334-344 (2000).

14  Dreyer, C. E., Janotti, A., Van de Walle, C. G. & Vanderbilt, D. *Phys. Rev. X.* **6**, 021038 (2016).

15  Bechstedt, F., Grossner, U. & Furthmüller, J. *Phys. Rev. B* **62**, 8003-8011 (2000).

16  Bernardini, F., Fiorentini, V. & Vanderbilt, D. *Phys. Rev. B* **63**, 193201 (2001).

17  Benbedra, A., Meskine, S., Boukortt, A., Abbassa, H. & Abbes, E. *Ecs J Solid State Sc* **12**, 103008 (2023).